\documentclass[prb,twocolumn,showpacs,showkeys]{revtex4}
\usepackage{graphicx}
\usepackage{graphics}
\usepackage{dcolumn}
\usepackage{bm}
\newcommand{\beq}{\begin{equation}}
\newcommand{\eeq}{\end{equation}}
\newcommand{\beqnar}{\begin{eqnarray}}
\newcommand{\eeqnar}{\end{eqnarray}}
\newcommand{\bfig}{\begin{figure}}
\newcommand{\efig}{\end{figure}}
\pdfoutput=1
\begin{document}
\title{A Gate-Induced Switch in Zigzag Graphene Naoribbons and Charging Effects}

\author{Hosein Cheraghchi, Hanyieh Esmailzade}

\affiliation{School of Physics, Damghan University of Basic
Sciences, P. O. Box: 36715- 364, Damghan, IRAN }
\email{cheraghchi@dubs.ac.ir}
\date{\today}

\begin{abstract}
Using non-equilibrium Green's function formalism, we investigate
nonlinear transport and charging effects of gated graphene
nanoribbons (GNRs) with even number of zigzag chains. We find a
negative differential resistance (NDR) over a wide range of gate
voltages with on/off ratio $\sim 10^6$ for narrow enough ribbons.
This NDR originates from the parity selection rule and also
prohibition of transport between discontinues energy bands. Since
the external field is well screened close to the contacts, the
NDR is robust against the electrostatic potential. However, for
voltages higher than the NDR threshold, due to charge transfer
through the edges of ZGNR, screening is reduced such that the
external potential can penetrate inside the ribbon giving rise to
smaller values of off current. Furthermore, on/off ratio of the
current depends on the aspect ratio of the length/width and also
edge impurity. Moreover, on/off ratio displays a power law
behavior as a function of ribbon length.

\end{abstract}
\pacs{73.23.-b,73.63.-b} \keywords {Graphene Nanoribbons, Negative
Differential Resistance, Electrostatic Potential Profile}
\maketitle
\section{Introduction}
Graphene is one of the intriguing new materials, which has been
studied extensively since Novoselov et al.\cite{novoselov}
fabricated it by micromechanical cleavage in $2004$. In fact,
flat structure of graphene makes its fabrication more
straightforward than carbon nanotubes. Moreover, dreams of carbon
nanoelectronic approach to the reality based on planar graphene
structures. This structure overcomes some difficulties of
nanoelectronics based on carbon nanotubes, by using lithography,
one-dimensional ribbon patterns on graphene sheets
\cite{experiment}. For achieving realistic nanoelectronic
applications based on graphene nanoribbons (GNR), width of ribbon
have to be narrow enough that a transport gap is
opened\cite{han,subtennm}. Sub-10 nm GNR field-effect-transistors
with smooth edges have been obtained in
Ref.[\onlinecite{subtennm}] and demonstrated to be semiconductors
with band-gap inversely proportional to the width and on/off
ratio of current up to $10^6$ at room temperature.

The origin of transport gap which is opened in a gate voltage
region of suppressed nonlinear conductance is still not well
understood\cite{Molitor,Louie,Castro-neto}. Based on the
tight-binding approach, GNRs with armchair shaped edges are
either metal or semiconductor\cite{Nakada,Brey,Louie,Zheng}.
Moreover, in this approach, zigzag edge ribbons are metal
regardless of their widths\cite{Onipko}. While {\it ab initio}
calculations\cite{Louie} predict that regardless of the shape of
the edges, GNRs are semiconductor. Two factors are responsible
for transport gap: the edge disorder leading to
localization\cite{Lewenkopf} and the
confinement\cite{Nakada,Brey,Zheng,Onipko}. However, in nonlinear
regime, transport gap is also opened by transition selection
rules which originates from the reflection symmetry\cite{Duan}.

Similar to carbon nanotubes, electronic transition through a
ZGNRs follows from some selection rules. The rotational symmetry
of the incoming electron wave function with respect to the tube
axis is conserved while passing through nanotubes\cite{farajian}.
Correspondingly, the transverse reflection symmetry of the
incoming and outgoing wave functions results in the parity
conservation in ZGNRs with even number of zigzag
chains\cite{grosso,beenaker,P-Ngraphene,duan,kurihara,Wakabayashi}.
As a consequence of the even-odd effect, a negative differential
resistance (NDR) region appears in the I-V characteristic curve
of P-N even ZGNR junctions\cite{P-Ngraphene}. A similar NDR
behavior has been also reported in P-N nanotube
junctions\cite{farajian}. Historically, NDR was first observed in
the degenerated N-P diode junctions\cite{esaki}. Nowadays, NDR has
been reported in many other molecular
devices\cite{NDRgraphene,cheraghchi}.
\bfig
\includegraphics[width=9 cm]{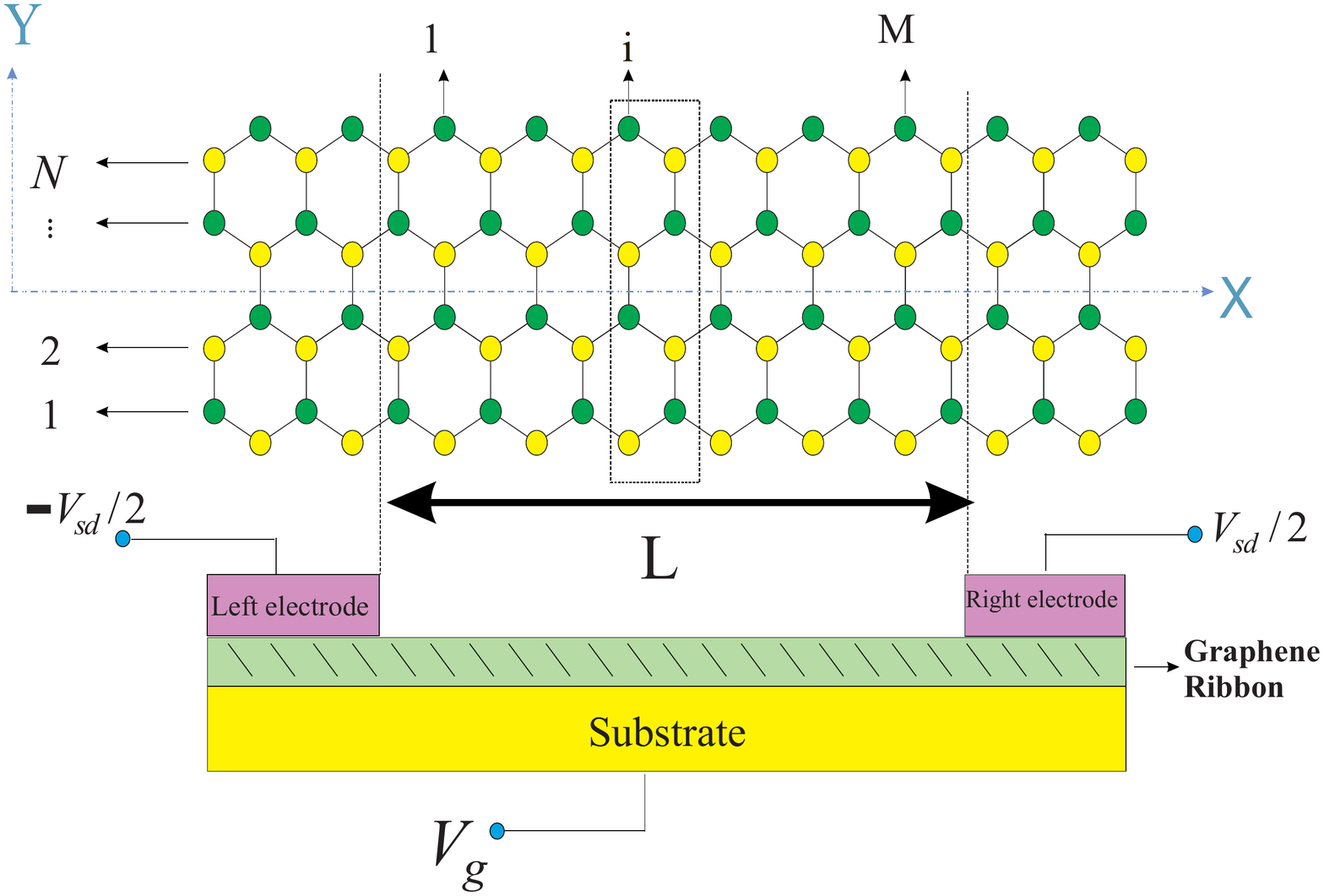}
\caption{Gated Zigzag graphene nanoribbon which is divided into
three regions: left, right and central region. Dotted rectangular
is the unit cell which is used for finding the band structure of
graphene ribbons. Lower panel shows a field-effect transistor
structure based on graphene ribbons where the gate voltage is
applied on the whole system. }\label{nanoribbon} \efig

Motivated by the work done in Ref.[\onlinecite{subtennm}], in this
paper, we investigate nonlinear transport through gated even
ZGNRs by using non-equilibrium Green's function (NEGF) approach.
The NDR region with on/off ratio of the current up to $10^6$
appears in the current-voltage characteristic curve. This NDR is
induced by transport gaps which are opened by two selection rules
governing electron transition through ZGNRs: (i) the parity
conservation, and (ii) that allowed transition are between
connected bands\cite{grosso}. Based on band structure analyzing,
we show that transport gap opened by the second selection rule is
filled for ribbons wider than $10nm$. So, sub-$10nm$ ribbons with
long enough length provide experimental manifestation of the NDR
phenomenon in I-V curve of GNRs. On the other hand, the gate
voltage regulates the current flow by shifting the blocked energy
regions with respect to the Fermi level. Moreover, on/off ratio
of the current displays a power law behavior as a function of
ribbon length as $M^{7.5}$.

Our calculations show that the details of the electrostatic
potential profile along the ribbon can not affect the emergence
of NDR. The same conclusion has been reported by
Ref.[\onlinecite{P-Ngraphene}], but they have not elaborated on
the physical reason behind this robustness. By following the
self-consistent charge and potential profiles at different
voltages, we demonstrate that at low voltages, strong screening
of the external potential at contacts results in a flat
electrostatic potential along the ribbon. Subsequently, the e-e
interaction at a mean field level, does not change the magnitude
of $I_{\rm on}$. However, for voltages higher than the NDR
threshold $V_{\rm on}$, the transfer of charge along the edges,
leads to more reduction in $I_{\rm off}$ which improves the switch
performance.

This paper is organized as follows: although the formalism has
been presented elsewhere\cite{cheraghchi}, to be self-contained,
we briefly introduce the Hamiltonian and NEGF formalism in section
II. In section III, we discuss selection rules governing even
ZGNRs. The origin of NDR seen in the I-V curve is explained in
section IV. We demonstrate in section V that the e-e interaction
does not have a significant effect on the phenomena of NDR in the
I-V curve. The last section concludes our results.
\section{Hamiltonian and Formalism}
Fig.(\ref{nanoribbon}) shows schematic side view of graphene
nanoribbon. In presence of source-drain applied potential, ribbon
is divided into three regions; left, right electrodes and also
central interacting region. Gate voltage is applied by means of
substrate on the graphene plate. The interacting Hamiltonian
governing the electron dynamics is written in tight-binding
approximation. This Hamiltonian is a functional of charge density:
\bfig
\includegraphics[width=9 cm]{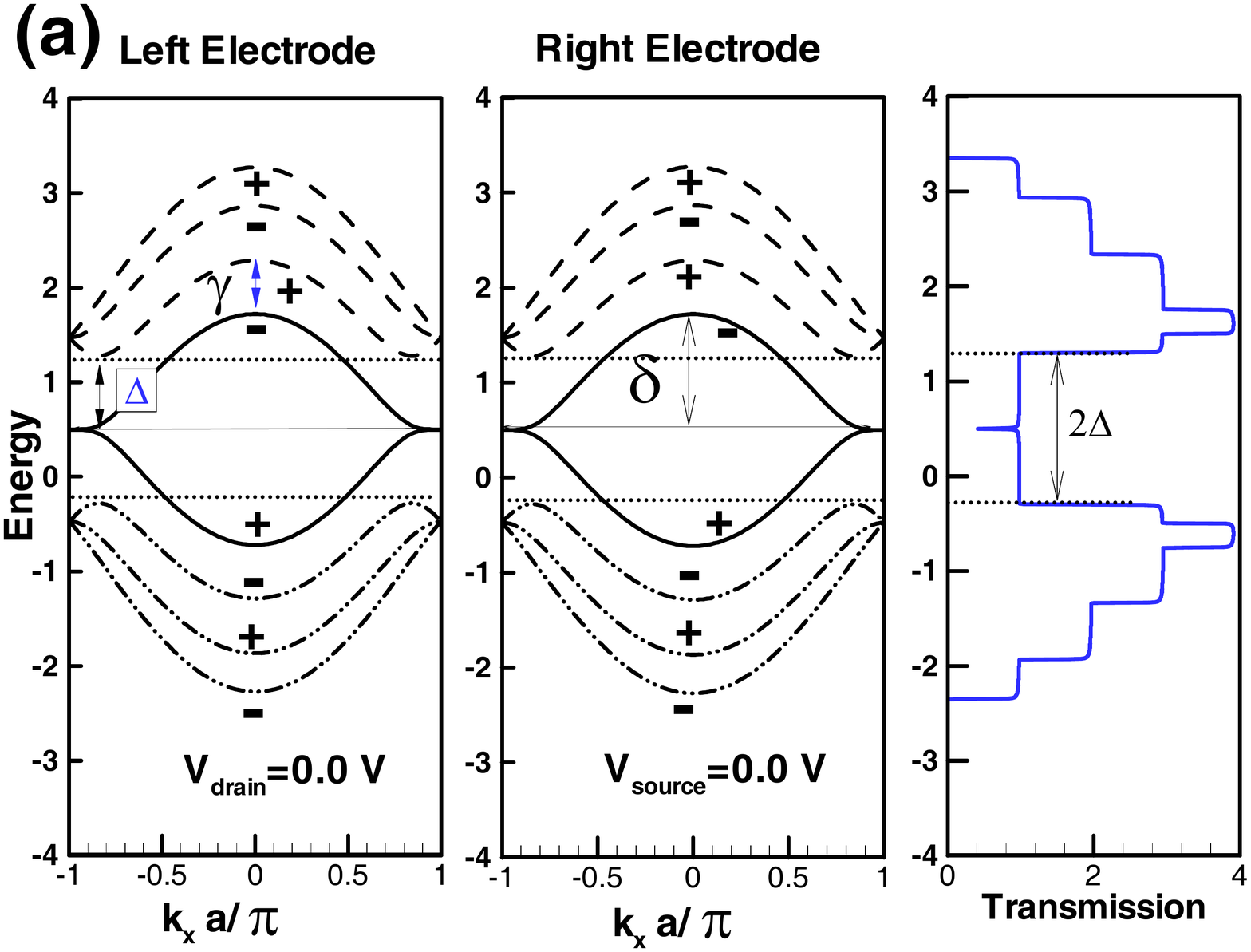}
\includegraphics[width=9 cm]{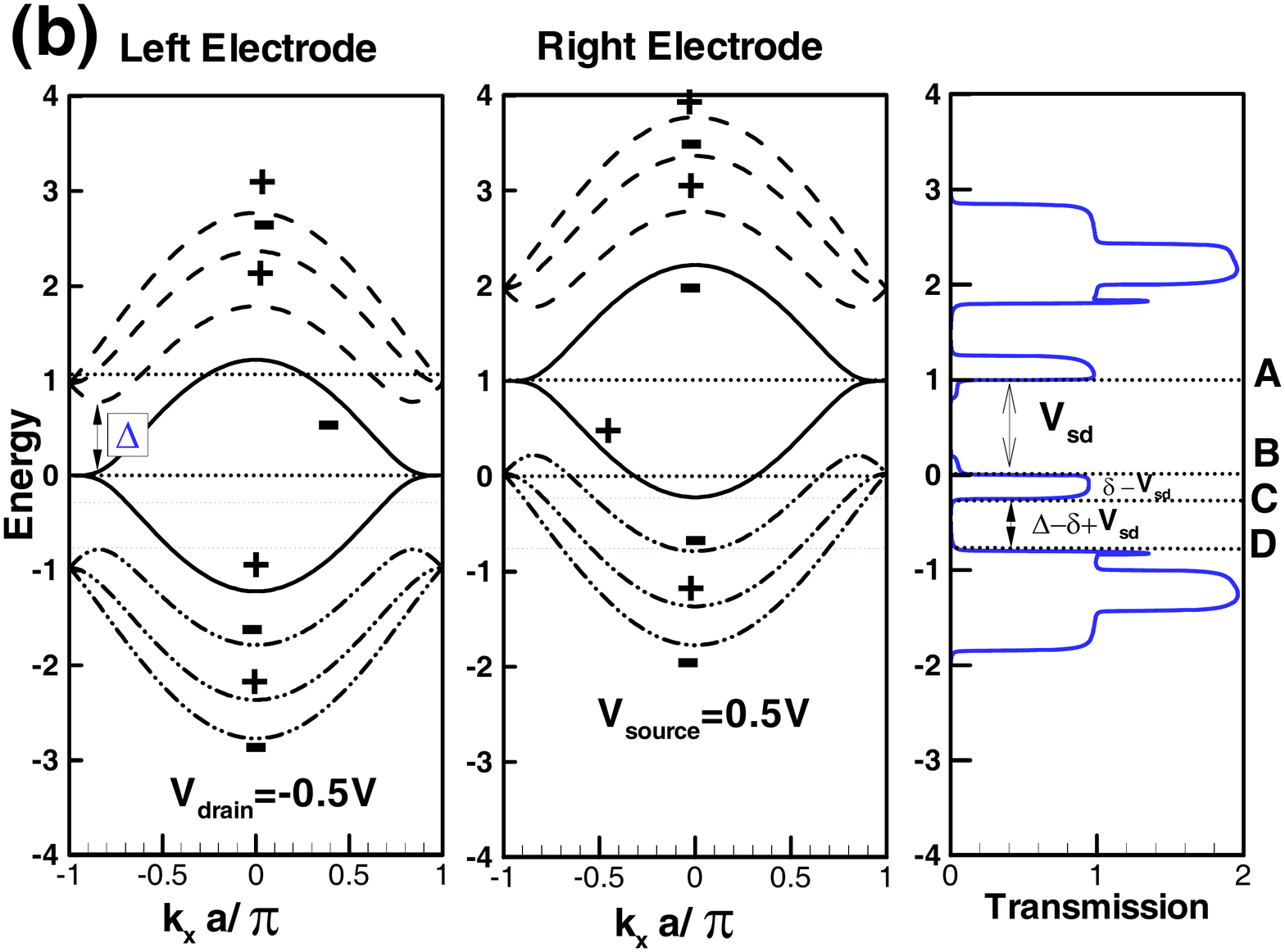}
\caption{transmission (right panel) and band structure of right
(center panel) and left (left panel) electrodes for ZGNR with $6$
unit cells in length and $4$ zigzag chains. Applied bias is
considered to be {\bf a)} $V_{\rm sd}=0$ and {\bf b)} $V_{\rm
sd}=1.0$. Gate voltage is $V_g=0.5 V$. Here, $\triangle$ and
$\gamma$ are the energy separation of upper/lower group of bands
from the central bands at the Dirac point ($k_x=\pm 2\pi/3a$) and
$k_x=0$, respectively. Moreover, $\delta$ is the half-width of
the central bands at $k_x=0$. }\label{spectrum} \efig

\beq\begin{array}{r}
H\{n\}=\sum_{i}(\varepsilon_i+[(x_i-x_0)/L-0.5]V_{sd}+\sum_{j}
U_{ij}\delta n_j]) c^{\dagger}_{i}c_{i}
\\ +\sum_{<ij>}t (c^{\dagger}_{i}c_{j}+c_{i}c^{\dagger}_{j}),
 \end{array} \label{hamiltonian}\eeq
$\varepsilon_i$ shows onsite energy of $i^{th}$ carbon atom and
$t$ represents the hopping integral between nearest neighbor
atoms. One $\pi$ orbital is considered per each site for graphene
system. Without losing any generality, we set onsite energies
($\varepsilon_i$) of all sites equal to zero. All energies are in
units of $t_{\rm C-C}=2.5 eV$. Application of a gate voltage is
achieved by shifting atomic onsite energies in all three regions.
The applied source-drain potential, $V_{\rm sd}$, and the gate
voltage, $V_{\rm g}$, preserve transverse symmetry with respect to
the ribbon axis ($X$ direction in Fig.(\ref{nanoribbon})). Linear
variation of the source-drain voltage along the ribbon is the
solution of the Laplace equation with Dirichlete boundary
condition on the contacts. $U_{ij}$ is the electrostatic Green's
function and $\delta n_i=n_i-n_i^0$ is the change in the
self-consistent charge $n_i$ from its initial equilibrium
zero-bias value. This third term is the direct Coulomb interaction
created by the bias-induced charges at a mean field level which
is the solution of Poisson equation. The electrostatic Green's
function for a distribution of charges between two parallel
conducting planes located at $x=0,L$ which are held at zero
potential\cite{Jackson}, has the following form:

 \beq\begin{array}{c}
U(\overrightarrow{r},\overrightarrow{r}^{'})=2\int_0^{\infty} dk
J_0(\alpha k)\frac{\sinh(k z_<)\sinh(k (L-z_>))}{\sinh(k L)} ,\\ \\
\alpha=\sqrt{(x-x^{'})^2+(y-y^{'})^2 + U_H^{-2}},\end{array}
 \label{exact}\eeq
where $U_H$ is the Hubbard parameter whose semi-empirical value
for carbon\cite{sctb} is about $4t_{\rm C-C}$. This parameter
determines the strength of electron-electron interaction. This
electrostatic Green's function is appropriate for the kernel of
Ohno-Klopmann model\cite{ok}.

First of all, we must find the self-consistent charge and
electrostatic potential by using NEGF formalism. The retarded and
advanced Green's function matrix subjected to the central portion
of the ribbon is as the follows: \beq
G^{r,a}(E,n)=[(E\pm\eta)I-H\{n\}-\Sigma_L^{r,a}-\Sigma_R^{r,a}]^{-1},
\label{Green}\eeq where $\eta\rightarrow0^+$. "$I$" is the unit
matrix. $\Sigma_{L/R}^r$ are the retarded self-energies due to
scattering by the left/right electrodes. These self-energies are
independent of the charge density. To determine the self-energy,
one needs to calculate the surface Green's function of
semi-infinite electrodes $g_{p}(E)$ by using the Lopez-Sancho's
method\cite{nardelli}. The escaping rate $\Gamma$ of electrons to
the electrodes is related to the self-energies as
$\Gamma_p=i[\Sigma_p^r-\Sigma_p^a] $. Having Green's function,
one can find total charge ($n=n^{eq}+n^{non-eq}$) by separate
calculations of the equilibrium and non-equilibrium charges by
using the retarded and lesser Green's functions, respectively. It
is simply demonstrated that in the coherent regime, lesser
Green's function can be represented by the retarded and advanced
Green's functions which are determined in Eq.(\ref{Green}),
  \beq\begin{array}{c}
n^{eq}_i=\frac{-1}{\pi}\int_{-\infty}^{\mu_0-V_{sd}/2}Im[G^r_{ii}(E)]dE
,\\ \\
n^{non-eq}_i=\frac{1}{2\pi}\int_{\mu_0-V_{sd}/2}^{\mu_0+V_{sd}/2}[G^r(\Gamma_L
f_L+\Gamma_R f_R)G^a]_{ii}dE, \end{array}\label{charge} \eeq
where $f_p=1/[1+\exp(\frac{E-\mu_p}{k_B T})]$ shows the Fermi
function of the electrodes, and $\mu_0=\mu_R=\mu_L$ show
electrochemical potentials of the left and right electrodes. The
initial charge $n_i^0$ is calculated by the equilibrium
integration in zero bias. To obtain the charge, coupled equations
of \ref{hamiltonian} and \ref{Green} are self-consistently solved
by using Broyden's method\cite{broyden-ohno}. The transmission
coefficient $T(E,V)$ is defined in terms of self-consistent
Green's functions as:
 \beq T={\rm Tr}[G^r \Gamma_R G^a \Gamma_L].\label{transmission}\eeq

The current passing through nanoribbon is calculated by the
Landauer formula at zero temperature \cite{Liang}:
 \beqnar
I(V)=\frac{2e}{h}\int_{\mu_0-V_{sd}/2}^{\mu_0+V_{sd}/2}\,dE
\,T(E,V)\ \label{current}.\eeqnar

\section{Band Structure and Selection Rules}
To understand the transport properties of ZGNRs, one needs to
study the band structure along with transmission curves. Band
structure of a wide ZGNR possesses two well-separated valleys
(conically-shaped dispersion curves) at $K$ and ${K}^{'}$ points
which can be effectively described by $2+1$ Dirac theory. Edge
states give rise to flat bands at charge neutrality point. The
band structure of ZGNR is calculated in tight-binding
approximation. The eigenvalues and eigenfunctions of the system
are extracted by diagonalization of the following
Hamiltonian\cite{Ezawa}:
$H=H_0+\tau_{0,+1}e^{ik_xx}+\tau_{0,-1}e^{-ik_xx}$; where $H_0$ is
the tight-binding Hamiltonian for the unit cell shown in
Fig.(\ref{nanoribbon}). In the nearest neighbor approximation,
$\tau_{0,+1} , \tau_{0,-1}$ are overlap matrices between a given
cell ('0') and its right, left neighboring cells, respectively.
These matrices are $2N\times2N$ dimensions where $N$ is the
number of zigzag chains. The eigenvalues $E(m,k_x)$ of this
Hamiltonian, give the energy spectrum of graphene ribbon shown in
Fig.(\ref{spectrum}). Here, $m$ and $k_x$ represent the band index
and longitudinal wave vector, respectively.
\bfig
\includegraphics[width=9 cm]{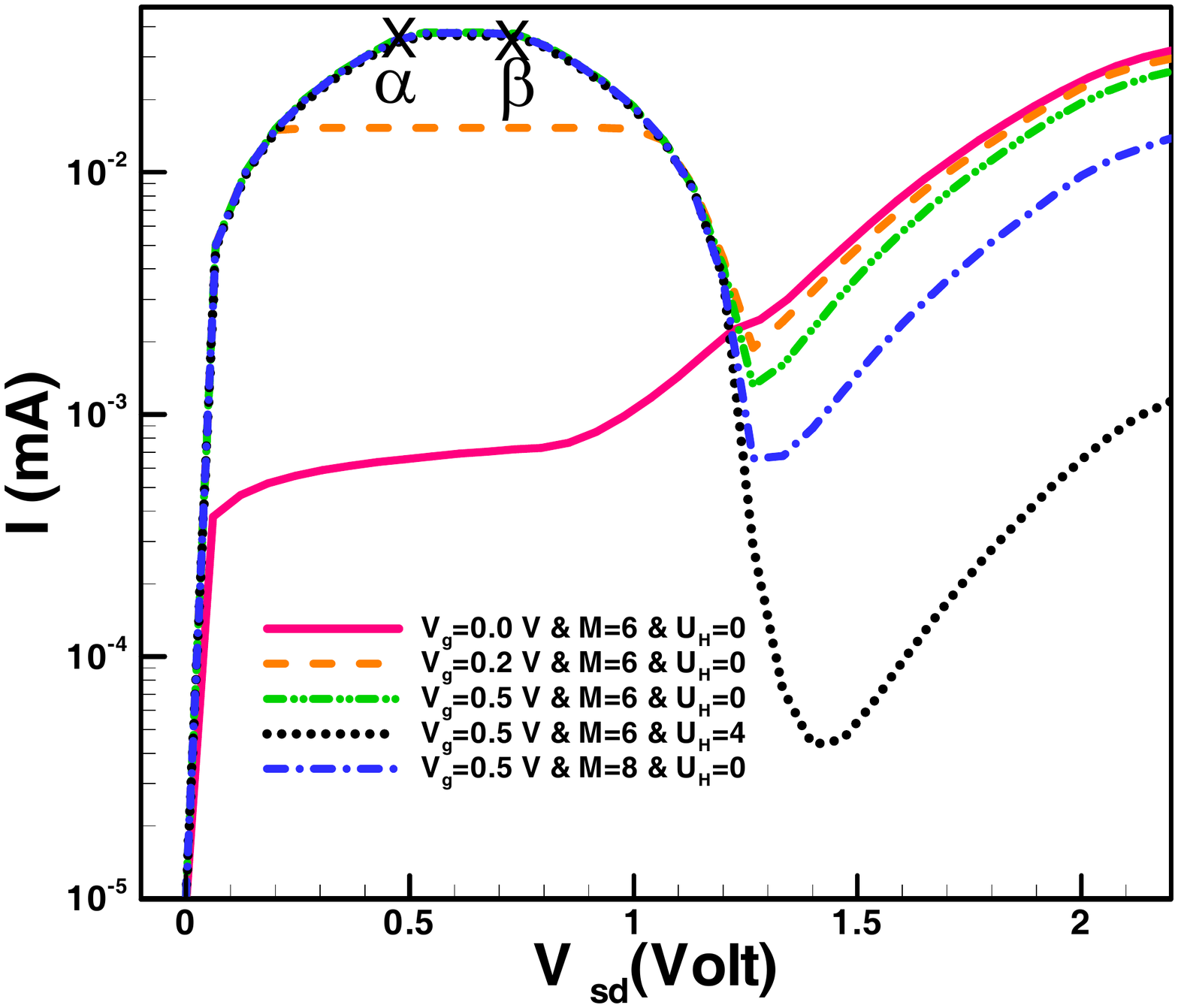}
\caption{Current-voltage characteristic curve for different gate
voltages when the ribbon size is as $(N,M)=(4,6)$. The effect of
different parameters such as size effect, electrostatic potential
and also gate voltage is investigated on I-V curve. Hubbard
parameter ($U_H$) is on site Coulomb repulsive.}\label{IV} \efig
The eigenfunction can be symmetric (even parity) or asymmetric
(odd parity) with respect to the mirror symmetry axis
\cite{Polini}. As a result of this mirror symmetry, all
eigenfunctions of each band have either '+' or '-' parity, which
is indicated in Fig.(\ref{spectrum}). If $N$ is even, the lowest
conduction band has even parity and the highest valance band has
odd parity. If the bands are labeled by the number $m$ from
bottom to top as $m=1,2,..,2N$, parity of $m^{th}$ band is
$(-1)^{(m+1)}$. In ZGNR, we consider longitudinal source-drain
potentials $V_{\rm sd}$ which are invariant under mirror symmetry.
Therefore, such $V_{\rm sd}$, does not destroy the parity of the
energy bands, although it changes the longitudinal momentum of
electron during transport.

Electronic transition is controlled by two selection rules. First,
the parity is conserved in tunneling of electron through
even-ZGNRs. Therefore, at zero source-drain voltage, one can
expect full transmission which is shown in
Fig.(\ref{spectrum}.a). In this case, all bands with the same
parity are energetically aligned and there is no gap in the
transmission curve. In Fig.(\ref{spectrum}.a), the energy of
transmission curve is shifted by the gate voltage ($0.5 V$).
Parity of each band is indicated by plus/minus signs. In the
range of $2 \Delta$, there is one conducting channel which
results in unit transmission coefficient. Transmission curve at
$V_{\rm sd}=0$ could also be extracted from
Fig.(\ref{CPtransmission}) which shows a contour plot of
transmission in plane of energy and $V_{\rm sd}$.

Fig.(\ref{spectrum}.b) represents band structures of electrodes
which are shifted with respect to each other due to the
source-drain voltage $V_{\rm sd}=1.0 V$. The gap in the
transmission curve of Fig.(\ref{spectrum}.b), AB region, indicates
that transport between bands of opposite parity is blocked.

The second selection rule which governs electron transport
through ZGNRs, is that, electron transition is allowed between
connected bands. Figs.(\ref{spectrum}.a,b) show the band
structure classified in three different groups; namely, central,
upper and lower bands which are indicated by solid, dashed and
dashed-dot-dot lines, respectively. The common feature of bands
in each group is that, they are connected at the zone boundary,
while distinct groups are disconnected. When one considers the
electron transport, the longitudinal momentum $k_x$, of electrons
changes as a result of applied $V_{\rm sd}$. The precise form of
this variation in $k_x$, crucially depends on profile of the
superimposed longitudinal potential. These groups are
disconnected from each other from the point of longitudinal
momentum. Variation of momentum of electron $k_x$ depends on the
shape of superimposed longitudinal potential. The transport
properties for smoothly varying $V_{\rm sd}$, are significantly
different from $V_{\rm sd}$ profiles with sharp spatial
variations. The electronic transition between an eigenstate
$(m_1,k)$ in the right electrode and an eigenstate $(m_2,q)$ in
the left electrode is proportional to Fourier transform of
longitudinal voltage and structure factor\cite{grosso},
\beq
 \langle\psi_{m_1}(k)\mid V_{sd}(x) \mid \psi_{m_2}(q)\rangle= S
 \tilde{V}_{sd}(k-q),
 \eeq
 where structure factor of $S$ is equal to
$[1+(-1)^{P_{m_1}+P_{m_2}}]$ for even ZGNRs and parity of band $m$
is equal to $P_{m}=(-1)^{(m+1)}$. Parity selection rule in even
ZGNRs originates from this structure factor. This parity
selection rule is mesoscopic analogue of chirality factors
governing transport of Dirac electrons in planar
graphene\cite{kirczenow}.

In our calculations, we apply a constant gate voltage to the
whole system without any longitudinal variation. As a result, the
gate voltage does not change momentum of electron.
\bfig
\includegraphics[width=9 cm]{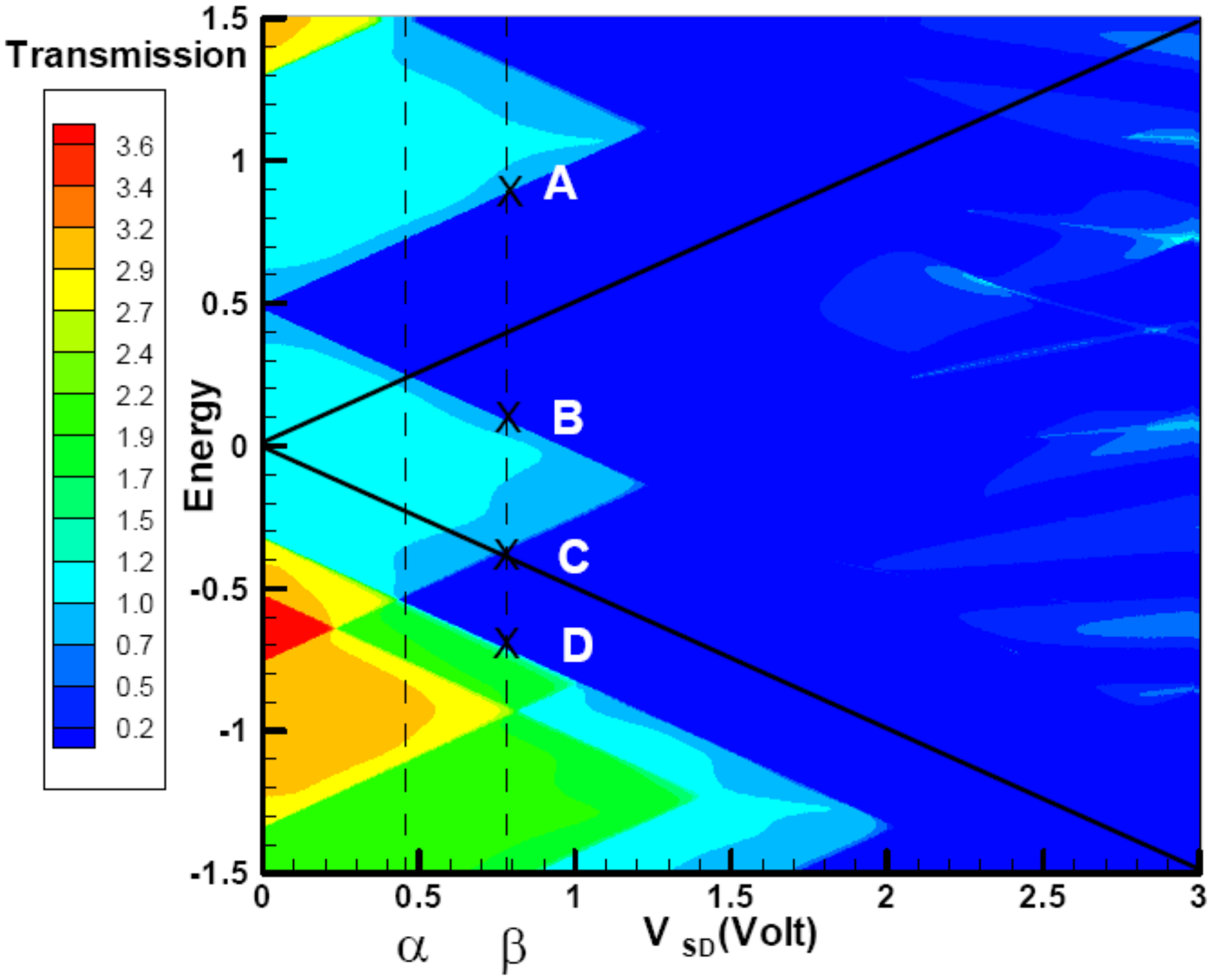}
\caption{ Contour plot of transmission with respect to the energy
and $V_{\rm sd}$ for the system size (N,M)=(4,6) and the applied
gate voltage $V_g=0.5 V$ . Darked Oblique lines shows the current
integration window. Points marked by $A, B, C, D$ correspond to
the horizontal lines with the same name in Fig.(\ref{spectrum}).
Lines of $\alpha ,\beta$ are the trace of points with similar
names in Fig.(\ref{IV}). Fermi energy was fixed at zero $E_F=0$.
}\label{CPtransmission} \efig
However, linear variation of the applied source-drain bias (with
the slope $V_{\rm sd}/L$) changes the electron momentum. So,
smooth variation of the potential in longer ribbons results in a
small momentum variation of electron. Consequently, transition of
electron between disconnected bands is forbidden when the length
of ribbon is so large that one can assume
$\tilde{V}_{sd}(k-q)\rightarrow\delta(k-q)$. Therefore, a smooth
potential in the longitudinal direction can just scatter the
electron among the class of states belonging to the same group of
connected energy bands.

Now let us focus on the two transport gap regions: AB and CD in
Fig.(\ref{spectrum}.b). The AB gap is a consequence of parity
selection rules, while the CD gap is due to blockage of
transition between disconnected groups. As can be seen in
Fig.(\ref{spectrum}.b), the AB gap is proportional to the
source-drain voltage, $V_{sd}$. Moreover, this gap is independent
of the ribbon width. Of course, in wide ribbons, upper and lower
band groups approach to the central group, especially at point
$k_x=0$, where $\gamma$ in Fig.(\ref{spectrum}.a) tends to zero
as log-normal.

When the ribbon width is increased, the separation $\gamma$
between the upper/lower and central groups of bands, is reduced,
which tends to loosen the second selection rule based on band
groups; hence filling in the gaps. However when we increase the
ribbon length, our classification of bands into connected groups
is recovered. Therefore the AB gap is essentially governed by the
aspect ratio of ZGNR.

The CD gap is equal to $\Delta-\delta+V_{\rm sd}$, where $\Delta$
and $\delta$ are the energy separation of upper/lower group from
the central bands at the Dirac point, and the half width of the
central bands at $k_x=0$, respectively. The dependence of
$\Delta$ on width N is: $\Delta\propto
(2.13\pm0.02)N^{-(0.864\pm0.003)}$, while $\delta$ has a
Log-Normal behavior which asymptotically approaches to the
constant value of $0.9738\pm0.0002$ as $N$ goes up to $10$. The
conducting region BC in Fig.(\ref{spectrum}.b) can exist only,
when $\delta-\Delta<V_{sd}<\delta$. From the dependence of
$\Delta$ and $\delta$ on $N$, the CD gap exists if $N$ is less
than $30$. Hence, NDR is estimated to be observable for ribbon
width $\leq 7 nm$. The lowest achieved ribbon width is
sub-$10nm$-wide ($\sim 2\pm0.5 nm$)\cite{subtennm} with the
length $\sim 236 nm$. Such long ribbons with small width provide
fascinating experimental manifestation of the selection rules in
transport properties.

\section{Negative Differential Resistance}

Fig.(\ref{IV}) shows current-voltage characteristic curve of a
ZGNR with $4$ zigzag chains and $6$ unit cells in length. In the
case of zero gate voltage, flow of current is blocked due to the
parity selection rule, while at a given $V_{\rm sd}$, gate bias
turns the current on. After a range of $V_{\rm sd}$ in which the
current remains unchanged, current begins to reduce with
increasing $V_{\rm sd}$. NDR threshold voltage $V_{\rm on}$
decreases with gate voltage for $V_g<0.6 V$. Dependence of NDR
threshold voltage on the gate voltage can also be seen in
Fig.(\ref{3Dcurrent}). This NDR also symmetrically appears in the
negative polarity of $V_{\rm sd}$. The NDR threshold voltage and
$I_{\rm on}$ remain unchanged in the presence of the
electron-electron interaction (with a given Hubbard term
$U=4t_{C-C}$). However, reduction of the current in off state,
$I_{\rm off}$, is intensified when one takes electrostatic
potential into account.
\bfig
\includegraphics[width=9 cm]{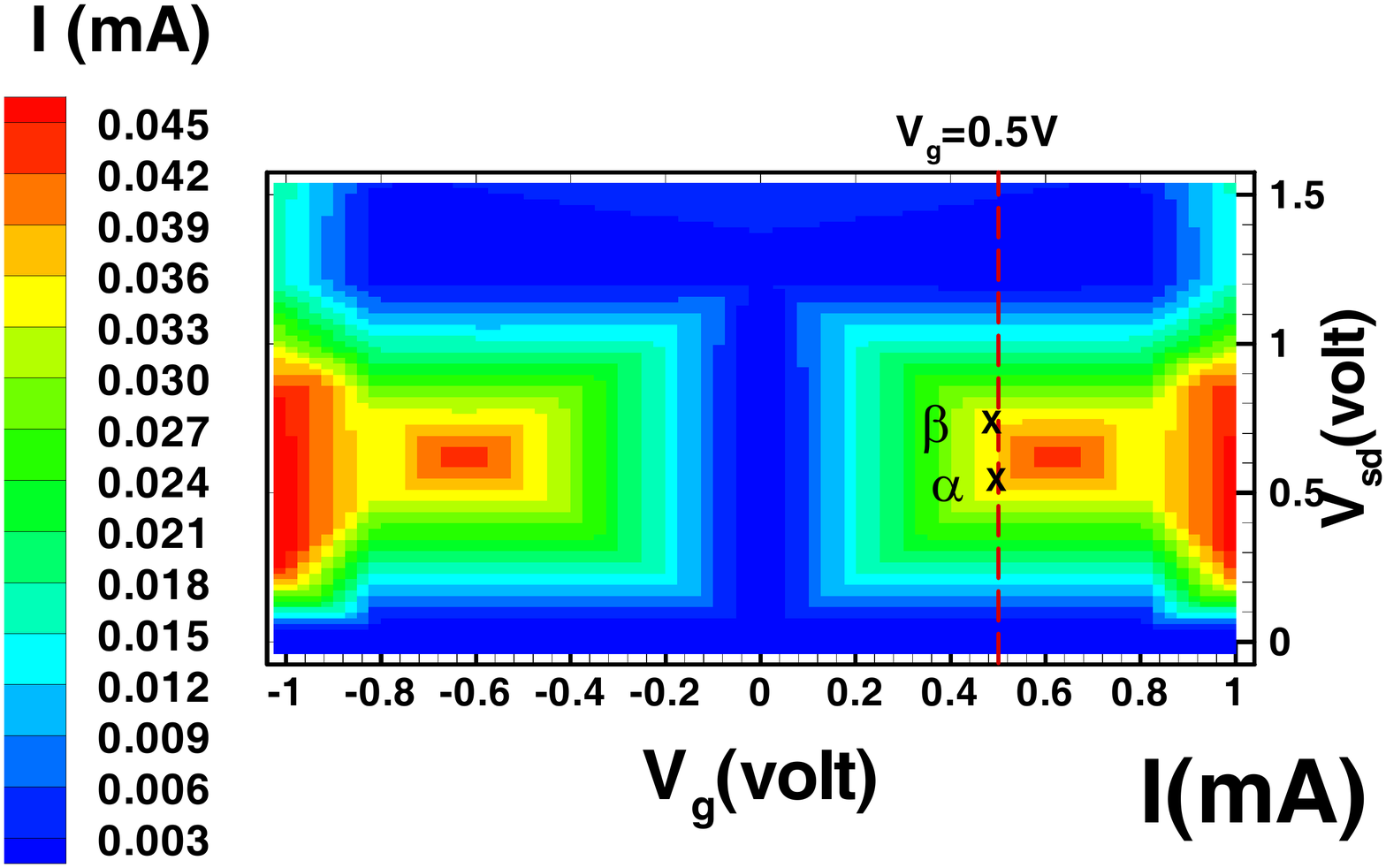}
\caption{Contour plot of the current in terms of $V_{sd}$ and
$V_g$ for a zigzag graphene nanoribbon with (N,M)=(4,6). Vertical
dotted line corresponds to $V_g=0.5 V$. $\alpha$ and $\beta$ are
those points which was shown in
Figs(\ref{IV},\ref{CPtransmission}). }\label{3Dcurrent} \efig

To understand the origin of NDR, it is helpful to look at the 3D
contour-plot of transmission in plane of energy and $V_{\rm sd}$
which is presented in Fig.(\ref{CPtransmission}). Blocked energy
intervals AB and CD, which are indicated in
Fig.(\ref{CPtransmission}), correspond to those intervals shown
in Fig.(\ref{spectrum}). For voltages lower than the vertical
line $\alpha$ ($V<V_{\alpha}$), transmission is a nonzero constant
for the whole region of the conduction window represented in
Eq.(\ref{current}). As a result, current increases proportionally
to $V_{\rm sd}$. In the voltage interval $[V_{\alpha},V_{\beta}]$,
the blocked region AB, originating from the parity selection rule,
contributes to the current integration window of
Eq.(\ref{current}). However, nonzero range of transmission
remains unchanged along with $V_{\rm sd}$ resulting in the fixed
current in the voltage range $[V_{\alpha},V_{\beta}]$. So, current
remains unchanged in this range. For voltages $V\geq V_{\beta}$,
the CD gap contributes in the current integration window, and
consequently the NDR phenomenon emerges.

Regarding the importance of gate voltage in the current flow, let
us investigate the effect of the gate voltage on $I-V_{\rm sd}$
curve by contour plotting of the current with respect to $V_{\rm
sd}$ and $V_g$ in Fig.(\ref{CPtransmission}). For gate voltages
$|V_g|<0.1V$, shift of transmission is not remarkable enough to
contribute to conducting channels in the current integration. So,
current is blocked by the parity selection rule. In the range
$0.1V<|V_g|<0.6V$, contribution of conducting region BC in
transport is accompanied with the blockage arising from AB and CD
gaps in voltages $V>V_{\rm on}$. As a consequence, current
reduces after a threshold voltage. In this range, on/off ratio of
the current increases and $V_{\rm on}$ reduces with increasing
the gate voltage.
\bfig
\includegraphics[width=9 cm]{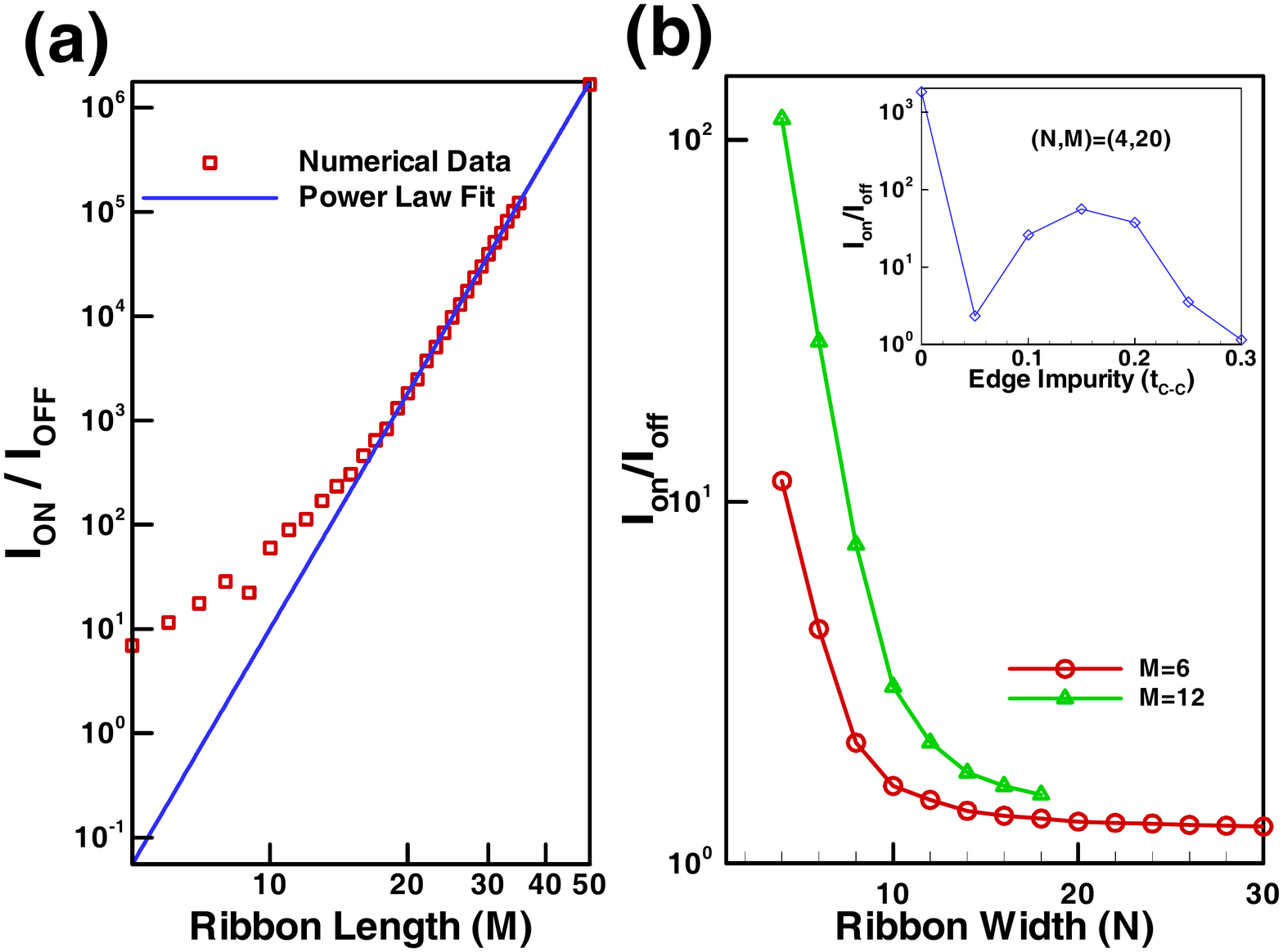}
\caption{ a) On-off ratio of the current increases as a power law
with the ribbon length for $N=4$ and $V_g=0.5 V$. b) On-off ratio
of the current decreases with the ribbon width (circle and
triangular points) while it is disappeared for edge impurity
(diamond points) stronger than 0.3$t_{C-C}$ (inset figure).
}\label{transmissionIonIoff} \efig

As can be seen in I-V curves of Fig.(\ref{IV}), off current
reduces for longer ribbons which enables us to achieve high
performance switches by increasing aspect ratio of the ribbon
length/width. The reason is connected to smooth variation of the
applied potential along the ribbon such that during transport,
electrons are scattered among those states which belong to
continuous bands. As a consequence, blockage originating from
electronic transition between disconnected bands is intensified
by increasing the ribbon length. In fact, when the length of
ribbon increases, transmission in the AB and CD gaps decreases
exponentially.

Since off current is induced by contribution of the gaps in the
current integration, $I_{\rm off}$ efficiently decreases with
increasing the ribbon length ($M$).
Fig.(\ref{transmissionIonIoff}.a) shows that $I_{\rm on}/I_{\rm
off}$ displays a power law behavior as a function of the ribbon
length for large $M$: $I_{\rm on}/I_{\rm off}\propto M^\eta$ where
$\eta=7.5061\pm0.03505$. As an example, for $M=50$, on/off ratio
goes up to $10^6$ which suggests experimental fabrication of high
performance switches based on the GNR nanoelectronics.

Experimentally, it was observed in Ref.[\onlinecite{subtennm}]
that the room-temperature on/off ratio induced by the gate voltage
increases exponentially as the GNR width decreases. They observed
that $I_{\rm on}/I_{\rm off}$ is equal to $1,5,100$ and $>10^5$
for $W=50nm,20nm,10nm$ and sub-$10nm$, respectively. Similarly, as
shown in Fig.(\ref{transmissionIonIoff}.b), on/off ratio
calculated for the set up considered in this paper, also decreases
with the ribbon width, while reduction of on/off ratio can be
compensated by considering longer ribbons. However, NDR
phenomenon is disappeared for the ribbons wider than $7nm$.

In {\it ab initio} calculations\cite{Louie}, by using
hydrogen-termination of zigzag edges, mirror-symmetry of ZGNRs
and consequently parity conservation could be retained.
Correspondingly, by several repetition of the heat treatments and
hydrogenation, it is also possible to create well-ordered
H-terminated edges in experiment\cite{kobayashi}. However, the
edge states with energies about $-0.1$ to $0.2$ eV have been
experimentally observed \cite{kobayashi} that emerge at
hydrogen-terminated zigzag edges. To simulate the edge states and
the effect of symmetry breaking on NDR phenomenon, it is assumed
to dope one of the ZGNR edges by slight impurity. Edge impurity
is considered to apply as a change in the on-site energy of the
edge atoms ($\varepsilon_\alpha$) with respect to on-site energy
of the other atoms. In case of edge disorder, $\varepsilon_\alpha$
plays the role of the averaged on-site energy of the edge atoms.
Inset figure indicated in Fig.(\ref{transmissionIonIoff}.b) shows
that on/off switching reduces with the edge impurity strength,
however, NDR still emerges for $\varepsilon_{\alpha} <
0.3t_{C-C}$.

\section{Electrostatic potential and charging effect on NDR}
Emerging phenomenon of negative differential resistance in I-V
curve is not destroyed by the e-e interaction and is independent
of the details of electrostatic potential profile. However,
interaction reduces off-current as shown in the I-V curves of
Fig.(\ref{IV}). To substantiate the above claim, comparison of
transmission curves in the presence and absence of the e-e
interaction is useful. It is apparent from
Fig.(\ref{t-int-nonint}.a) that for voltages less than $V_{\rm
on}$, transmission in conducting channels is robust against the
e-e interaction while transmission increases in the gaps with
respect to the non-interacting case. But this enhancement is
slight enough which can not affect the emergence of NDR. However,
for voltages $V>V_{\rm on}$, interaction lowers transmission
coefficient in the conducting channels (such as BC region) in
which higher subbands participate in transport. Such behavior is
corroborated in Fig.(\ref{t-int-nonint}.b) which indicates
transmission at that voltage corresponding to the off-current,
$V_{\rm off}$. Reduction in the transmission coefficient of the
conducting channels results in further reduction of the
off-current.
 \bfig
\includegraphics[width=9 cm]{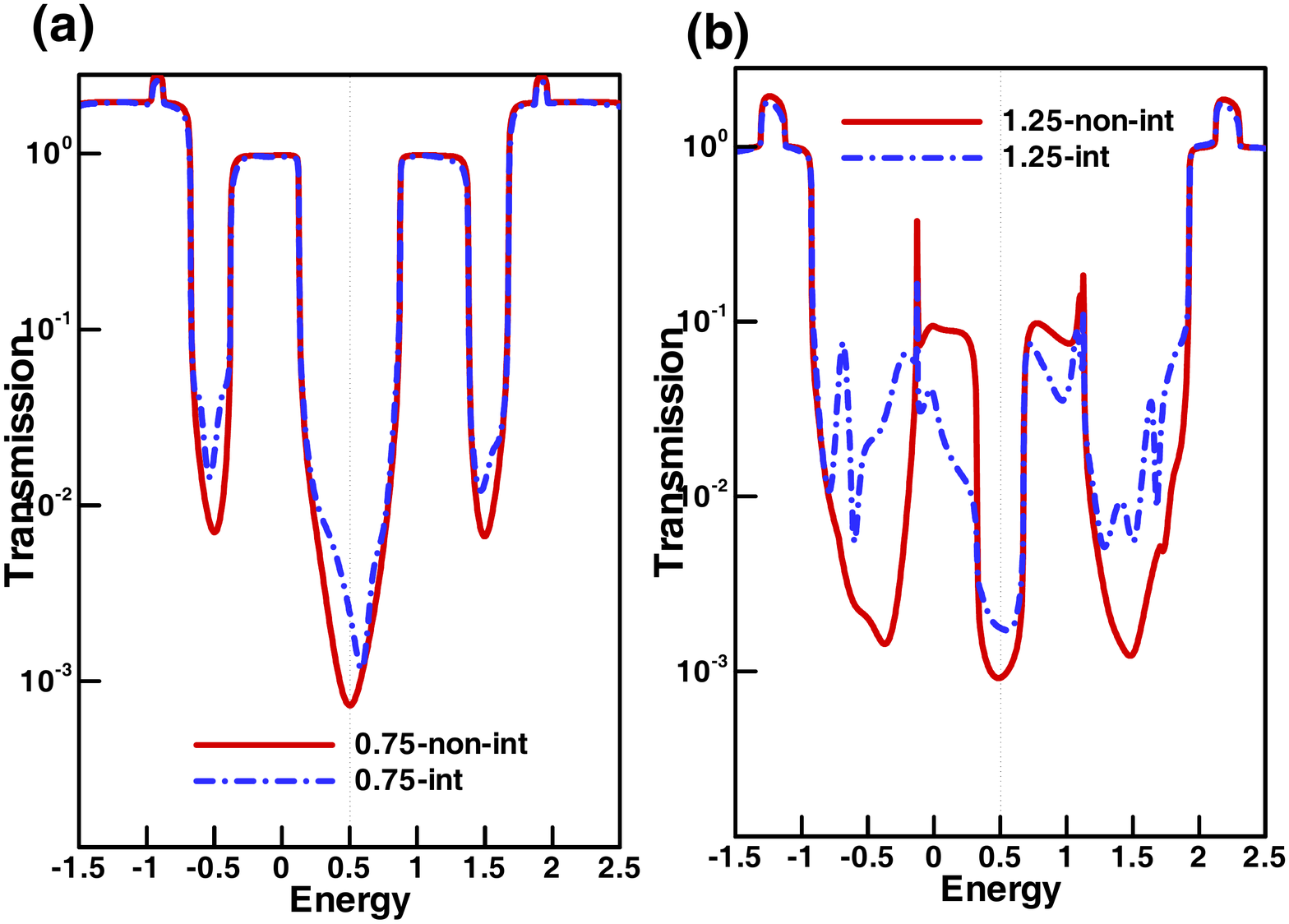}
\caption{ Transmission curve of zigzag graphene nanoribbon with 4
zigzag chains in width for voltages a) 0.75 volt b) 1.25 volt.
Interacting and non-interacting curves are compared with each
other.}\label{t-int-nonint} \efig
To explain the reason for such phenomenon, it is necessary to
study the potential and charge profiles. Electrostatic potential
averaged on each unit cell is represented in
Fig.(\ref{chg-pot-v}.a) in terms of the ribbon length. For
voltages less than $V_{\rm on}$, potential sharply drops only at
the contact regions which connects the system to electrodes. In
such a case, external potential is strongly screened by
redistributed electrons and, electrostatic potential of the
central atoms remains close to zero. Screening is performed by
discharging of electrons from the area connected to the source
and their accumulation around the drain electrode. These facts
are obvious from transferred charge and electrostatic potential
profiles represented in Fig.\ref{chg-pot-profile}. Since
$U(n-n_0)$ determines electrostatic behavior of the potential,
discharging of electrons weakens the external potential
penetrated from the source electrode. Moreover, charge
accumulation around the drain electrode prevents potential drop
in the central part of the system. So in the case of strong
screening, potential drops only at the contacts.
\bfig
\includegraphics[width=9 cm]{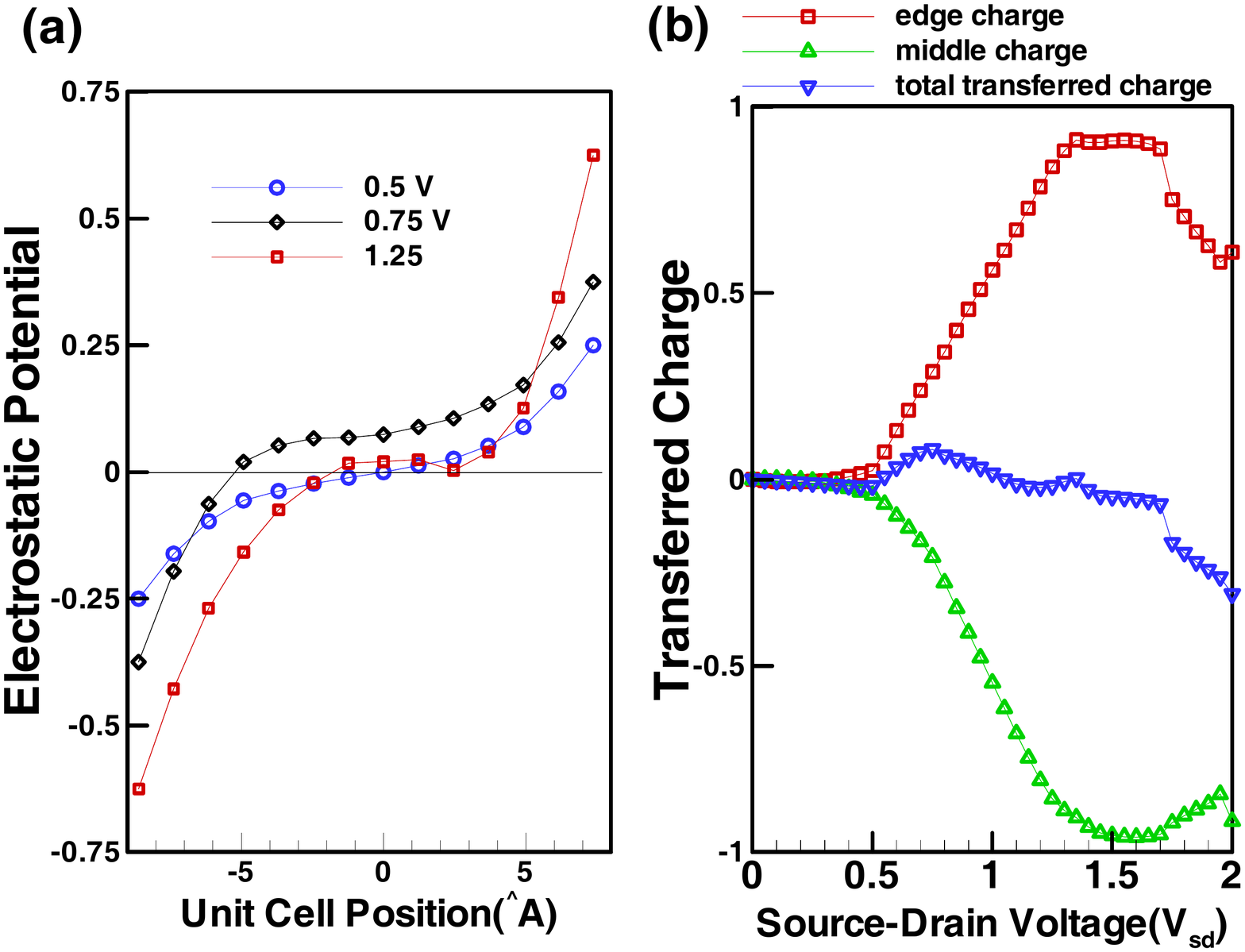}
\caption{a) Edge, middle and total transferred charge in terms of
source-drain applied bias. b) Electrostatic potential per unit
cell in terms of unit cell position for source-drain voltages 0.5,
0.75, 1.25 volts. The gate voltage is for all curves equal to
$V_g=0.5$ V. }\label{chg-pot-v} \efig
However, when the applied bias goes beyond $1$ volt, screening is
being weakened and external potential can penetrate inside the
central region. The reason as to why screening is weak, can be
sought in charge distribution. Figure (\ref{chg-pot-v}.b)
illustrates that for voltages less than $0.5$ volt, in and out
flow of charge are balanced with each other such that the total
transferred charge remains close to zero. However, around the
voltage $V_{\rm on}$ and voltages above, the charge is mainly
transferred from the edges of the ribbon, so that the source
electrode does not inject further charge to middle atoms of the
ribbon. As a consequence, by increasing the applied bias and so
gradient of the potential along the central region, charge
depletion is mainly enhanced in the middle bar area of ZGNR. On
the other hand, since the only way for transporting electrons is
the edge atoms, significant accumulation of charge appears along
the two edge lines of ZGNR.

In summary, at voltages less than $V_{\rm on}$, electrostatic
potential is only dropped at the contacts and therefore momentum
of electrons is only varying in the area where the potential
drops, while longitudinal momentum of electron remains unchanged
across the central portion. In other words, potential steeply
drops in the low-area district around the contacts which results
in violation of the blockage rule which governs on transition
between disconnected energy bands. Subsequently, transmission
coefficient slightly increases in the blocked energy ranges. In
other words, in this case, an increase in gradient of the
potential facilitates electronic transport in the blocked
energies. Note that interaction preserves transverse symmetry, so
the parity selection rule still governs electronic transport.
Therefore, the AB gap induced by the parity conservation still
survives for voltages larger than $V_{\rm on}$. For voltages
$V>V_{\rm on}$, electrostatic potential gradually penetrates into
the whole system so that the potential of the central region is
not flat. In addition, because the edge transport of electrons
dominates, the transverse potential is deeper in the middle of
ZGNR than its edges. Therefore, the band structure of the
interacting central region differs from the band structure of
electrodes. As a consequence, for voltages $V>V_{\rm on}$,
transmission of conducting channels and also off current reduces.
\bfig
\includegraphics[width=9 cm]{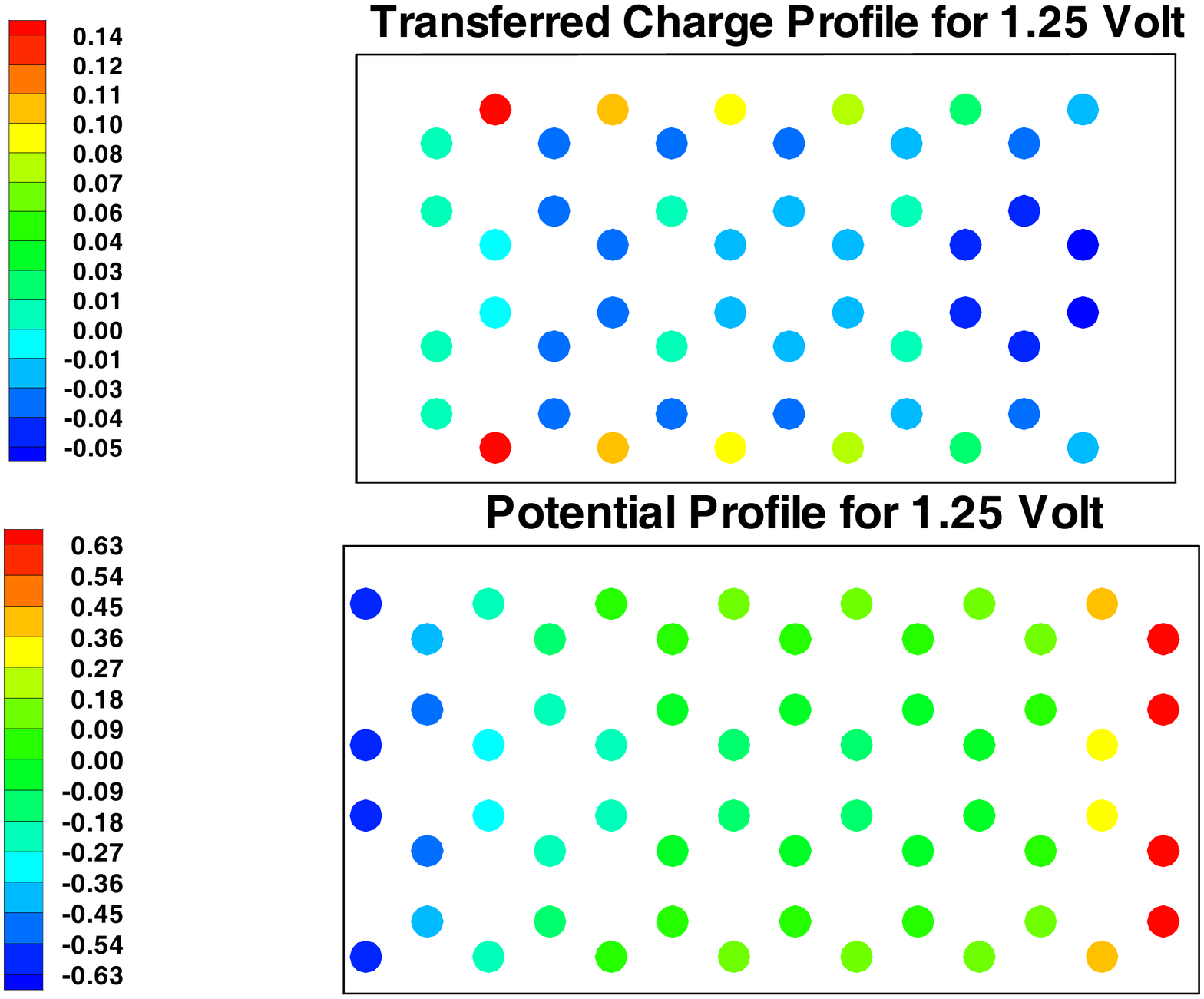}
\caption{Transferred charge and Electrostatic potential profiles
for $V_{\rm sd}=1.25$ volt and $V_g=0.5$ volt in the weak
screening case. Due to charge transfer through the ribbon edges,
screening is so weak that the external potential can penetrate
inside the central portion. }\label{chg-pot-profile} \efig

\section{Conclusion}
In this paper, using non-equilibrium Green's function approach,
we investigated nonlinear transport and charging effects of
graphene nanoribbons with even zigzag chains. Current flow is
controlled by the gate voltage applied on the whole of sub-$10
nm$ ribbons. In this range of widths, two selection rules govern
on the electronic transition are: (i) the parity conservation, and
(ii) allowed transition between connected bands. As a result, a
negative differential resistance (NDR) in I-V curve is appeared in
the presence of the gate voltage. Furthermore, on/off ratio of
the current increases with the ribbon length as a power law
behavior up to $10^6$, while ribbon width and edge impurity reduce
on/off ratio. Emergence of the NDR phenomenon is not sensitive to
details of the electrostatic potential profile. Because of strong
screening in low voltages, the major potential drop takes place
at the contacts. However, in voltages larger than the NDR
threshold, due to charge transfer through the ribbon edges,
screening is so weak that the external potential can penetrate
inside the central portion. As a consequence, off current reduces
in comparison to non-interacting ribbons.

\section{Acknowledges}
We wish to acknowledge Prof. K. Esfarjani for
collaboration\cite{cheraghchi} in the development of NEGF
formalism and Dr. S. A. Jafari for a critical proof-reading of
the manuscript and for his useful comments and suggestions.

\end{document}